\documentclass{article}
\usepackage[margin=2cm]{geometry}

\usepackage[T1]{fontenc}
\usepackage{hyperref}
\usepackage{url}
\usepackage{booktabs}
\usepackage{babel}
\usepackage{amsfonts}
\usepackage{nicefrac}
\usepackage{microtype}
\usepackage{xcolor}
\usepackage{amsthm}
\usepackage{amsmath}
\usepackage{graphicx}
\usepackage{natbib}
\usepackage{authblk}
\title{Latent Spaces for Langevin Dynamics}

\date{}

\author[1]{
  Andy Bruce\thanks{acbruce@ucsc.edu}
}
\author[1]{
  Alexander Aghili\thanks{awaghili@ucsc.edu}
}
\author[1]{
  Razvan Marinescu\thanks{ramarine@ucsc.edu}
}
\author[1]{
  Daniel Sabo\thanks{dsabo@ucsc.edu}
}

\affil[1]{University of California, Santa Cruz}

\begin{document}

\maketitle

\begin{abstract}
  In the field of machine learning coarse-grained potentials in molecular dynamics, many propagators require that the effective Hamiltonian is quadratic in momentum, thus limiting the family of coarse-graining functions. In this paper, we derive a general family of coarse-graining embedding functions for which Langevin dynamics samples correctly. These equations have significant implications for molecular simulations and pave the way for Langevin dynamics on non-geometric coarse-graining representations, such as those provided by principal components of component analysis or latent embeddings of molecules obtained from neural networks.
\end{abstract}

\section{Introduction}
In molecular dynamics, coarse-graining (CG) speeds up simulations by mapping groups of atoms onto fewer effective degrees of freedom, thereby reducing the computational cost of integration. The CG function used determines which degrees of freedom are kept. For example, \cite{Husic2020} only keeps the Cartesian coordinates of the Carbon-alpha atom in each amino acid. Another common method is MARTINI (\cite{Marrink2007Martini}), which groups roughly four non-hydrogen atoms into a single coarse-grained bead. We consider a fine-grained (FG) all-atom system as given by all-atom molecular dynamics. The FG system has microstates defined as positions and momentum \(\vec{q}, \vec{p} \in \mathbb{R}^n\) and Hamiltonian \(H\). We aim to find a coarse-grained system that has macrostates with positions and momentum \(\vec{Q}, \vec{P} \in \mathbb{R}^N\) with the mapping \(\vec{Q} = f(\vec{q})\) for some latent space embedding function \(f \in \mathbb{R}^n \rightarrow \mathbb{R}^N\). In the canonical (NVT) ensemble, the effective CG Hamiltonian (free energy) for a macrostate as per \cite{gibbs1902elementary} should satisfy:
\[
F(\vec{Q}, \vec{P}) = -k_B T \ln(Z(\vec{Q}, \vec{P}))
\]
where \(Z\) is the constrained partition function. The gradient of the free energy should give the potential of mean force \cite{PhysRevE.105.054138}. Here we define \(Z\) to be
\begin{equation}
  \label{partition_func}
  Z(\vec{Q}, \vec{P}) = \frac{1}{h^n}\int d\vec{q} \int d\vec{p} e^{-\frac{H(\vec{q}, \vec{p})}{k_B T}}
  \prod_i^N \delta\Big(\vec{Q}_i - f_i(\vec{q})\Big) \prod_i^N\delta\Big(\frac{\partial F}{\partial P_i} - \sum_j^n \frac{\partial f_i(\vec{q})}{\partial q_j}\frac{\partial H}{\partial p_j}\Big)
\end{equation}
where  \(h\) is Plank's constant. The first Dirac delta term constrains the positions of FG microstates to match the CG macrostate, while the second term similarly constrains the velocities. Many bottom up CG derivations such as \citet{Noid2008} constrain the embedding so the atoms \(j \in I_i\) contributing weights \(a_{ij}\) to interaction site \(i\) are disjoint from all other interaction sites, so \(I_\alpha \cap I_\beta = \emptyset\) if \(\alpha \neq \beta\). Equivelently, it is a CG function that is just a straightforward linear transform \(f(\vec{q}) = \Xi \vec{q}\) where \(\Xi \in \mathbb{R}^{N \times n}\), where \(\Xi_{ij} = a_{ij}\), and the rows must be disjoint (so if \(\Xi_{ij} \neq 0\) then \(\Xi_{kj} = 0\) forall \(k \neq i\)). In this case, assuming a FG Hamiltonian of the form
\begin{equation}
  \label{hamiltonian_boring}
  H(\vec{q}, \vec{p}) = \sum_j \frac{p_j^2}{2m_{\text{FG}, j}} + U(\vec{q})
\end{equation}
the free energy, up to a constant, can be written in the form below.
\begin{equation}
  \label{cg_hamiltonian_boring}
F(\vec{Q}, \vec{P}) = \sum_i \frac{P_i^2}{2m_{\text{CG}, i}} + V(\vec{Q})
\end{equation}
where the effective CG masses are:
\begin{equation}
  \label{noid_masses}
  m_{\text{CG}, i} = \Big( \sum_{j \in I_i} \frac{a_{ij}^2}{m_{\text{FG}, j}} \Big)^{-1}
\end{equation}
In this work, we claim that we can perform CG dynamics with a more general form of the Hamiltonian:
\begin{equation}
  \label{hamiltonian_quadratic}
  H(\vec{q}, \vec{p}) = \frac{1}{2}\vec{p}^\top M^{-1}(\vec{q}) \vec{p} + U(\vec{q})
\end{equation}
where \(M\) is a symmetric matrix. In addition, we show that for some choices \(f\), the free energy can be written in the form below:
\begin{equation}
  \label{free_energy_quadratic}
F(\vec{Q}, \vec{P}) = \frac{1}{2}\vec{P}^\top R^{-1}(\vec{Q}) \vec{P} + V(\vec{Q})
\end{equation}
as long as there is a function \(R(\vec{Q})\) for the effective CG masses, satisfying
\begin{equation}
  \label{cg_mass_matrix}
  R^{-1}(f(\vec{q})) = J_f(\vec{q}) M^{-1}(\vec{q}) J_f^\top(\vec{q})
\end{equation}
where \(J_f\) is the Jacobian of \(f\). Here, \(R\) must only depend on the codomain of \(f\), implying \(J_f(\vec{q}) M^{-1}(\vec{q}) J_f^\top (\vec{q})\) is the same for all microstates \(\vec{q}\) in the preimage of \(\vec{Q}\), imposing a complex constraint on which \(f\)'s can be used. We show some general solutions and candidate \(f\)'s that satisfy these constraints in section \ref{candidates}.

In the previous case when \(f(\vec{q}) = \Xi \vec{q}\) with constant diagonal masses \(M\), it reduces to \(R^{-1} = \Xi M^{-1} \Xi^\top\) recovering Eq \ref{noid_masses}. However, unlike previous methods requiring disjoint rows of \(\Xi\), this method works for arbitrary \(\Xi\) as long as \(R\) is invertible. If the rows of \(\Xi\) are not disjoint, then there will be off-diagonal terms in \(R\), and the free energy will not be in the form of Eq. \ref{cg_hamiltonian_boring}.

\section{Free Energy Potential and Momentum Contributions}
With the FG Hamiltonian in Eq. \ref{hamiltonian_quadratic}, substituting it in Eq. \ref{partition_func} and making some minor assumptions, we analytically find an expression for the free energy:
\begin{multline}
  \label{big_equation}
  e^{-\frac{F(\vec{Q}, \vec{P})}{k_B T}} = \exp\Bigg(-\frac{1}{k_B T}\Big(\frac{1}{2}\vec{P}^\top R^{-1}(\vec{Q}) \vec{P} + V(\vec{Q})\Big)\Bigg) = \\
  \frac{O}{h^n}\int d\vec{q} \exp \Bigg( -\frac{1}{k_B T} \bigg(U(\vec{q}) -k_B T \ln\Big[\frac{C}{(2\pi)^N s^N O} \sqrt{\frac{(k_B T)^{-N} (2 \pi)^N}{\text{det}\Big( J_f(\vec{q}) M^{-1}(\vec{q}) J_f^\top(\vec{q})\Big)}} \sqrt{\frac{(k_B T)^n (2\pi)^n}{\text{det} M^{-1}(\vec{q})}} \Big]  \\
  + \frac{1}{2}\vec{P}^\top (R^{-1}(\vec{Q}))^\top \Big( J_f(\vec{q}) M^{-1}(\vec{q}) J_f^\top(\vec{q})\Big)^{-1} R^{-1}(\vec{Q}) \vec{P} \bigg) \Bigg) \prod_i^N \delta\Big(Q_i - f_i(\vec{q})\Big)
\end{multline}
where \(C\) is the product of all the units of the \(Q_i\)'s, \(O\) is the product of all the units of the \(p\)'s, and \(s\) is the unit of time. One can see that if Eq. \ref{cg_mass_matrix} is satisfied, then \(V\) can be solved for as below:
\begin{multline}
  \label{cg_potential}
  V(\vec{Q}) = -k_B T \ln\Bigg(\frac{O}{h^n}\int d\vec{q} \exp \Bigg( -\frac{1}{k_B T} \bigg(U(\vec{q}) \\
  -k_B T \ln\Big(\frac{C}{(2\pi)^N s^N O} \sqrt{\frac{(k_B T)^{-N} (2 \pi)^N}{\text{det} R^{-1}(\vec{Q})}} \sqrt{\frac{(k_B T)^n (2\pi)^n}{\text{det} M^{-1}(\vec{q})}} \Big) \bigg) \Bigg) \prod_i^N \delta\Big(Q_i - f_i(\vec{q})\Big) \Bigg)
\end{multline}
because the Dirac delta function removes the contributions of any points where the \(R^{-1}(\vec{Q})\) does not cancel in the last line of the exponent of \ref{big_equation}, which can then be cancelled on both sides.
\section{Applications}
\subsection{Machine Learning}
With machine-learned CG force fields, this method provides a way to separate the free energy into momentum and potential parts for a general set of CG functions \(f\), rather than only the usual disjoint weighted sums of the Cartesian coordinates of atoms into interaction sites. Because Eq. \ref{free_energy_quadratic} analytically calculates the momentum contribution, the neural network will only need to learn the potential portion of the free energy \(V(\vec{Q})\). The contribution of the momentum becomes trivial given an expression for \(R^{-1}(\vec{Q})\). Learning the potential \(V\) may be done with force matching as per \cite{Noid2008}, or relative entropy matching as per \cite{Shell2008}.

\subsection{Langevin Dynamics}

If the free energy is in the form of Eq. \ref{free_energy_quadratic}, it is known that Langevin dynamics can sample from a quadratic Hamiltonian of Eq. \ref{hamiltonian_quadratic} with the SDE
\[
d\vec{q} = \nabla_p H(\vec{q}, \vec{p}) dt = M^{-1}(q) \vec{p} dt
\]
\[
d\vec{p} = -\nabla_q H(\vec{q}, \vec{p})dt - \gamma M^{-1}(\vec{q}) \vec{p}dt + \sqrt{2\gamma k_B T} d\vec{W}
\]
where the steady state is the Boltzmann distribution as per \cite{Leimkuhler2015}.
\[
\rho(\vec{q}, \vec{p}) \propto e^{-\frac{H(\vec{q}, \vec{p})}{k_B T}}
\]
The macrostates can then also be sampled correctly with the SDE by substituting \(H\) for \(F\) and \(M^{-1}(\vec{q})\) for \(R^{-1}(\vec{Q})\) in the Langevin equation. The SDE will correctly sample the macrostates proportional to the free energy so that \(\rho(\vec{Q}, \vec{P}) \propto e^{-\frac{F(\vec{Q}, \vec{P})}{k_B T}}\). The Langevin equation will provide both ``configurational'' and ``momentum'' consistency (\cite{Jin2022}). The potential of mean force \(\nabla_Q F(\vec{Q}, \vec{P})\) (\cite{Ciccotti2005}) is expressed as
\begin{equation}
    \label{blue_moon_eq}
    \nabla_Q F(\vec{Q}, \vec{P}) = \Big\langle B(\vec{q}) \nabla_{q} H(\vec{q}, \vec{p}) \Big \rangle_{f(\vec{q}) = \vec{Q}} - k_B T \Big\langle \nabla_q \cdot B(\vec{q}) \Big\rangle_{f(\vec{q}) = \vec{Q}}
\end{equation}
and due to the quadratic separation, it can be written as
\[
\nabla_Q F(\vec{Q}, \vec{P}) = \Big\langle B(\vec{q}) \nabla_{q} E(\vec{q}) \Big \rangle_{f(\vec{q}) = \vec{Q}} + \nabla_Q \Big( P^\top R^{-1}(\vec{Q}) P \Big) - k_B T \Big\langle \nabla_q \cdot B(\vec{q}) \Big\rangle_{f(\vec{q}) = \vec{Q}}
\]
where \(E\) is the exponent expression in Eq. \ref{cg_potential}, and \(B(\vec{q})\) is any matrix satisfying the below. 
\[
B(\vec{q}) J^\top_f(\vec{q}) = I
\]
One common choice is the left pseudoinverse of the Jacobian (if it exists), although other methods exist, such as those described in \cite{Krmer2023}. These results illustrate that Langevin dynamics can be performed directly in non-geometric, machine-learned coarse-grained coordinates while maintaining full thermodynamic correctness.

\subsection{Hyperparameter Matching}
Since we now have an exact analytical expression for the free energy, similar to force matching, we can also match hyperparameters as per \cite{duschatko2024thermodynamicallyinformedmultimodallearning}. For example, if the NN model generalizes over temperature, then the entropy \(S\) and thus internal energy \(U\) can also be matched. Using \(\vec{\Gamma} = (\vec{Q}, \vec{P})\)
\[
F(T, \vec{\Gamma}) = -k_B T \ln(Z(\vec{\Gamma})) = U(T, \vec{\Gamma}) - T S(T, \vec{\Gamma})
\]
then taking the partial derivative gives
\[
\frac{\partial F(T, \vec{\Gamma})}{\partial T} = -S(T, \vec{\Gamma})
\]
and thus the internal energy can be calculated as below.
\[
U(T, \vec{\Gamma}) = F(T, \vec{\Gamma}) + T \frac{\partial F(T, \vec{\Gamma})}{\partial T}
\]
If the training data has the total Hamiltonian, then it can be matched with ``internal energy matching'' using mean squared error in addition to force matching.
\[
L_{\text{energy}}(\theta) = \frac{1}{M}\sum_{i=0}^M \Big\lVert \Big(W(\theta, T_i, D_i) - T_i \frac{\partial W(\theta, T, D_i)}{\partial T}\Big) - H(D_i) \Big\rVert^2
\]
This method requires differentiating the model with respect to temperature.

\newpage
\clearpage
\bibliographystyle{plainnat}
\bibliography{refs}

\appendix

\section{Proof for Equation \ref{big_equation}}
\begin{proof}
The derivation requires that \(M\) is positive definite, and that \(J_f M^{-1} J_f^\top\) is also positive definite. If \(M\) is positive definite then \(J_f M^{-1} J_f^\top\) will be positive definite as long as \(J_f\) is full rank.

We start with \ref{partition_func} and substitute in the FG Hamiltonian.
\[
Z(\vec{Q}, \vec{P}) = \frac{1}{h^n}\int d\vec{q} \int d\vec{p} e^{-\frac{\frac{1}{2}\vec{p}^\top M^{-1}(\vec{q}) \vec{p} + U(\vec{q})}{k_B T}} \delta^N\Big(\vec{Q} - f(\vec{q})\Big) \prod_i^N\delta\Big(\frac{\partial F}{\partial P_i} - \sum_j^n \frac{\partial f_i(\vec{q})}{\partial q_j}\frac{\partial H}{\partial p_j}\Big)
\]
Move the terms only dependent on position out of the inner integral.
\[
\frac{1}{h^n}\int d\vec{q} e^{-\frac{U(\vec{q})}{k_B T}}  \delta^N\Big(\vec{Q} - f(\vec{q})\Big) \int d\vec{p} e^{-\frac{\frac{1}{2} \vec{p}^\top M^{-1}(\vec{q}) \vec{p}}{k_B T}} \prod_i^N\delta\Big(\frac{\partial F}{\partial P_i} - \sum_j^n \frac{\partial f_i(\vec{q})}{\partial q_j}\frac{\partial H}{\partial p_j}\Big)
\]
Re-express the Dirac delta.
\[
\frac{1}{h^n}\int d\vec{q} e^{-\frac{U(\vec{q})}{k_B T}} \delta^N\Big(\vec{Q} - f(\vec{q})\Big) \int d\vec{p} e^{-\frac{\frac{1}{2} \vec{p}^\top M^{-1}(\vec{q}) \vec{p}}{k_B T}} \delta^N\Big(\nabla_P F(\vec{Q}, \vec{P}) - J_f(\vec{q}) \nabla_p H(\vec{q}, \vec{p}) \Big)
\]
Assume \(R^{-1}\) is symmetric.
\[
\frac{1}{h^n}\int d\vec{q} e^{-\frac{U(\vec{q})}{k_B T}} \delta^N\Big(\vec{Q} - f(\vec{q})\Big) \int d\vec{p} e^{-\frac{\frac{1}{2} \vec{p}^\top M^{-1}(\vec{q}) \vec{p}}{k_B T}} \delta^N\Big(R^{-1}(\vec{Q})\vec{P} - J_f(\vec{q}) M^{-1}(\vec{q})\vec{p} \Big)
\]
Next, use the Fourier representation of the Dirac delta.
\small
\[
\frac{1}{h^n}\int d\vec{q} e^{-\frac{U(\vec{q})}{k_B T}} \delta^N\Big(\vec{Q} - f(\vec{q})\Big) \int d\vec{p} e^{-\frac{\frac{1}{2} \vec{p}^\top M^{-1}(\vec{q}) \vec{p}}{k_B T}} \frac{s^N}{(2\pi)^NC}\int d\vec{k} e^{i\vec{k}^\top\Big(R^{-1}(\vec{Q})\vec{P} - J_f(\vec{q}) M^{-1}(\vec{q})\vec{p} \Big)}
\]
Assuming the necessary conditions, swap the integrals and move the nondependent terms out of the inner integral.
\[
\frac{s^N}{(2\pi)^Nh^nC}\int d\vec{q} e^{-\frac{U(\vec{q})}{k_B T}} \delta^N\Big(\vec{Q} - f(\vec{q})\Big)  \int d\vec{k} e^{i\vec{k}^\top R^{-1}(\vec{Q})\vec{P}} \int d\vec{p} e^{-\frac{\frac{1}{2} \vec{p}^\top M^{-1}(\vec{q}) \vec{p}}{k_B T} - i\vec{k}^\top J_f(\vec{q}) M^{-1}(\vec{q})\vec{p}}
\]
Complete the squares in the exponent.
\begin{multline}
  \frac{s^N}{(2\pi)^Nh^nC}\int d\vec{q} e^{-\frac{U(\vec{q})}{k_B T}} \delta^N\Big(\vec{Q} - f(\vec{q})\Big)  \int d\vec{k} \\
  e^{i\vec{k}^\top R^{-1}(\vec{Q})\vec{P}} \int d\vec{p} e^{-\frac{1}{2} \Big(\vec{p} - ik_BTJ_f^\top \vec{k}\Big)^\top \frac{M^{-1}(\vec{q})}{k_B T} \Big( \vec{p} - ik_BTJ_f^\top \vec{k} \Big) - \frac{1}{2}k_B T \Big(J_f^\top \vec{k} \Big)^\top M^{-1}(\vec{q}) \Big(J_f^\top \vec{k} \Big)}
\end{multline}
Move the nondependent part out.
\begin{multline}
  \frac{s^N}{(2\pi)^Nh^nC}\int d\vec{q} e^{-\frac{U(\vec{q})}{k_B T}} \delta^N\Big(\vec{Q} - f(\vec{q})\Big)  \int d\vec{k} e^{i\vec{k}^\top R^{-1}(\vec{Q})\vec{P}- \frac{1}{2}k_B T \Big(J_f^\top \vec{k} \Big)^\top M^{-1}(\vec{q}) \Big(J_f^\top \vec{k} \Big)} \\
  \int d\vec{p} e^{-\frac{1}{2} \Big(\vec{p} - ik_BTJ_f^\top k\Big)^\top \frac{M^{-1}(\vec{q})}{k_B T} \Big( \vec{p} - ik_BTJ_f^\top k \Big)}
\end{multline}
Since we assumed \(M^{-1}\) is positive definite, then the right-most integral can be evaluated as a Gaussian integral. Gaussian integrals are invariant under translations, even if the translation is complex.
\begin{equation}
  \frac{s^N}{(2\pi)^Nh^nC}\int d\vec{q} e^{-\frac{U(\vec{q})}{k_B T}} \delta^N\Big(\vec{Q} - f(\vec{q})\Big) \sqrt{\frac{(k_B T)^n (2\pi)^n}{\text{det} M^{-1}(\vec{q})}}\int d\vec{k} e^{i\vec{k}^\top R^{-1}(\vec{Q})\vec{P}- \frac{1}{2}k_B T \Big(J_f^\top \vec{k} \Big)^\top M^{-1}(\vec{q}) \Big(J_f^\top \vec{k} \Big)} 
\end{equation}
Complete the squares again.
\tiny
\begin{multline}
  \frac{s^N}{(2\pi)^Nh^nC}\int d\vec{q} e^{-\frac{U(\vec{q})}{k_B T}} \delta^N\Big(\vec{Q} - f(\vec{q})\Big) \sqrt{\frac{(k_B T)^n (2\pi)^n}{\text{det} M^{-1}(\vec{q})}} e^{-\frac{1}{k_BT} \frac{1}{2} P^\top (R^{-1}(\vec{Q}))^\top \big(J_f(\vec{q}) M^{-1}(\vec{q})J_f^\top(\vec{q})\big)^{-1} R^{-1}(\vec{Q}) P} \int d\vec{k}\\
  e^{-\frac{1}{2}k_B T \big(\vec{k} + \frac{i}{k_B T}\big(J_f(\vec{q}) M^{-1}(\vec{q})J_f^\top(\vec{q})\big)^{-1} R^{-1}(\vec{Q}) P \big)^\top \big(J_f(\vec{q}) M^{-1}(\vec{q})J_f^\top(\vec{q})\big) \big(\vec{k} + \frac{i}{k_B T}\big(J_f(\vec{q}) M^{-1}(\vec{q})J_f^\top(\vec{q})\big)^{-1} R^{-1}(\vec{Q}) P \big)}
\end{multline}
\normalsize
And once again assuming \(\big(J_f(\vec{q}) M^{-1}(\vec{q})J_f^\top(\vec{q})\big)^{-1}\) is positive definite, evaluate the Gaussian integral.
\begin{multline}
  \frac{s^N}{(2\pi)^Nh^nC}\int d\vec{q} e^{-\frac{U(\vec{q})}{k_B T}} \delta^N\Big(\vec{Q} - f(\vec{q})\Big) \sqrt{\frac{(k_B T)^n (2\pi)^n}{\text{det} M^{-1}(\vec{q})}} \sqrt{\frac{(k_B T)^{-N} (2\pi)^N}{\text{det} \big(J_f(\vec{q}) M^{-1}(\vec{q})J_f^\top(\vec{q})\big)}} \\
  e^{-\frac{1}{k_BT} \frac{1}{2}P^\top (R^{-1}(\vec{Q}))^\top \big(J_f(\vec{q}) M^{-1}(\vec{q})J_f^\top(\vec{q})\big)^{-1} R^{-1}(\vec{Q}) P}
\end{multline}
\end{proof}

\section{Coarse Grain Function Candidates} \label{candidates}
\subsection{Isometric and Maximal Invariance} 
Define the set of isometries \(G\) that also commute with \(M\) to be all functions of \(g(\vec{q}) = R\vec{q} + b\) where \(R M R^\top = M\). If \(f\) is transitive on level sets, so that \(\forall \vec{q}_1\ \forall \vec{q}_2 \ \   f(\vec{q}_1) = f(\vec{q}_2) \rightarrow \exists (g \in G)\ s.t.\  \vec{q_1} = g(\vec{q_2})\), and \(f\) is invariant on the subset of \(G\) used within level sets, so that \(f = f \circ g\), then equation \ref{cg_mass_matrix} can be satisfied. We can show that \(J_f(\vec{q}_1) M J^\top_f(\vec{q}_1) = J_f(\vec{q}_2) M J^\top_f(\vec{q}_2)\) for any microstates \(\vec{q}_1, \vec{q}_2 \in f^{-1}(\vec{Q})\) in the preimage of the same macrostate as follows:

\begin{proof}
\[
  J_f(\vec{q}_1) M J^\top_f(\vec{q}_1)
\]
Existentially instantiate the \(g\) for these two elements of the level set, and since \(f\) is invariant on \(g\)
\[
= J_{f \circ g}(\vec{q}_1) M J^\top_{f \circ g}(\vec{q}_1)
\]
and the Jacobian of function composition is the composition of the Jacobians.
\[
J_f(g(\vec{q}_1)) J_g(\vec{q}_1) M J_g^\top(\vec{q}_1)J^\top_f(g(\vec{q}_1))
\]
\[
= J_f(g(\vec{q}_1)) R M R^T J^\top_f(g(\vec{q}_1))
\]
\[
= J_f(\vec{q}_2) M J^\top_f(\vec{q}_2)
\]
This proof shows the existence of an \(R(Q)\) for any choice of \(f\) that satisfies the conditions.
\end{proof}
\subsection{Distance Latent Space}
Consider a coarse-grained function \(f(\vec{q})\) that takes the distances between pairs of atoms. \(f\) is isometric invariant almost by definition, so an \(R\) should exist. We can find it by defining
\[
f_{ab}(\vec{q}) = \sqrt{(q_{ax} - q_{bx})^2 + (q_{ay} - q_{by})^2 + (q_{az} - q_{bz})^2}
\]
Then
\[
(J_f M J_f^\top)_{ab, cd} = \sum_{i \in [n]} m_i \frac{\partial f_{ab}}{\partial \vec{q}_i } \frac{\partial f_{cd}}{\partial \vec{q}_i }
\]
Note that if \(ab \cap cd\) do not share endpoints, then it is just zero since one of the derivatives will always be zero. If they do, then let \(b\) be the atom that they share.
\[
(J_f M J_f^\top)_{ab, bc} = \sum_{i \in [n]} m_i \frac{\partial f_{ab}}{\partial \vec{q}_i } \frac{\partial f_{bc}}{\partial \vec{q}_i }
\]
One can compute
\[
\frac{\partial f_{ab}}{\partial q_{ak}} = \frac{(q_{ak} - q_{bk})}{f_{ab}}
\]
If \(ab = bc\) then 
\[
(J_f M J_f^\top)_{ab, ab} = \sum_{k \in \{x, y, z\}} m_a \frac{\partial f_{ab}}{\partial \vec{q}_{ak} } \frac{\partial f_{ab}}{\partial \vec{q}_{ak}} + \sum_{k \in \{x, y, z\}} m_b \frac{\partial f_{ab}}{\partial \vec{q}_{bk} } \frac{\partial f_{ab}}{\partial \vec{q}_{bk}} = m_a + m_b
\]
\[
m_a \frac{f_{ab}^2}{f_{ab}^2} + m_b \frac{f_{bc}^2}{f_{bc}^2} = m_a + m_b
\]
otherwise
\[
(J_f M J_f^\top)_{ab, bc} = m_b \frac{\sum_{k \in \{x, y, z\}}\Big((q_{ak} - q_{bk})(q_{ck} - q_{bk})\Big)}{f_{ab}f_{bc}} = m_b \cos(\theta_{abc})
\]
Thus, if \(f\) provides enough edges so that the structure is "rigid", then the angles can be determined from the cosine rule, and a formula for \(R\) is trivial.

\subsubsection{Distance Latent Space Results}
We trained a simple feed-forward network on the 1st, 6th, and 10th Carbon-alpha atom of Chignolin (\href{https://www.rcsb.org/structure/1UAO}{1UAO}) at 300K. The input was two distances of the sixth atom to the other two, and the cosine of the angle. The coarse-grain function is:

\[
f(\vec{q}) = \begin{bmatrix}
\lVert \vec{q}_1 - \vec{q}_6 \rVert \\
\lVert \vec{q}_{10} - \vec{q}_6 \rVert \\
\frac{(\vec{q}_1 - \vec{q}_6) \cdot (\vec{q}_{10} - \vec{q}_6)}{\lVert \vec{q}_1 - \vec{q}_6 \rVert \lVert \vec{q}_{10} - \vec{q}_6 \rVert }
\end{bmatrix}
\]
The model outputs a single scalar, free energy. The gradient of the free energy was then trained using force matching \cite{Noid2008}.

\[
L(\theta) = \frac{1}{M} \sum_{i=1}^M \Big \lVert \nabla_Q W(f(D_i), \theta) - \nabla_Q F(f(D_i)) \Big \rVert ^2
\]

where \(\nabla_Q F\) is the expression in Eq. \ref{blue_moon_eq}.

\begin{figure}[ht]
  \begin{center}
    \includegraphics[trim={5cm 7.3cm 5cm 8.5cm},clip,width=0.6\textwidth]{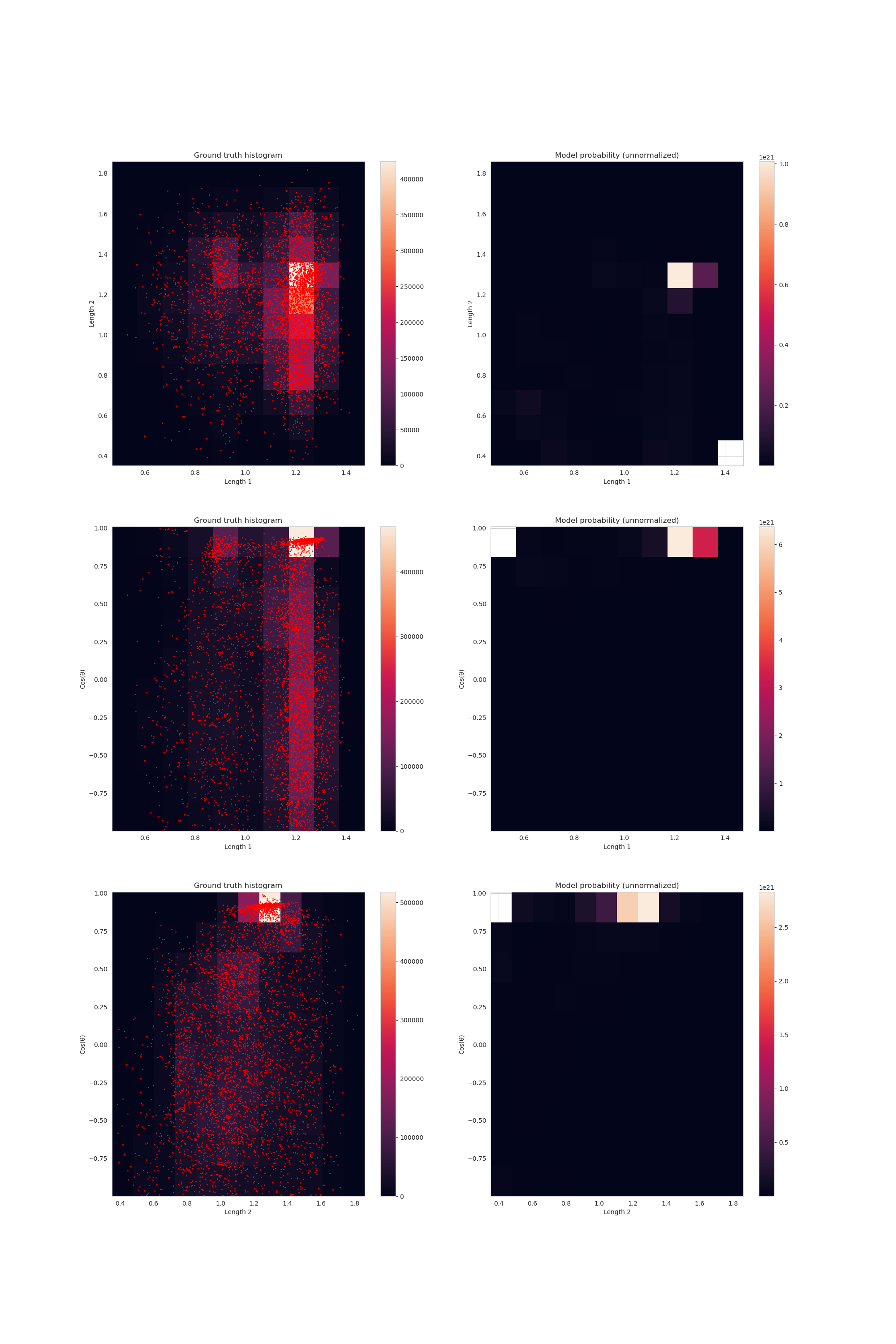}
  \end{center}
  \caption{Ground truth and model prediction histogram of the probabilities in latent space. Units of length are in nanometers. Red points are the strided positions of the ground truth data. The unnormalized model probability was evaluated using the Boltzmann distribution.}
\end{figure}
\end{document}